\documentclass[12pt,thmsa]{article}
\usepackage{amssymb}

\usepackage{sw20aip}

\input{tcilatex}
\begin{document}

\author{James L. Carr\thanks{%
Present address: 1509 Corcoran St. NW, Washington, D.C. 20009, U.S.A.,
e-mail:\ jimcarr@carrastro.com}}
\title{A classical picture of lepton neutral current forces}
\date{6 March 1999}
\maketitle

\begin{abstract}
When charged current weak interations are excluded, the neutral current weak
interaction is formally similar to ordinary electromagnetism with a massive
photon. In this spirit, the Maxwell equations for the fields of the Z-boson
are derived from the standard model. These describe the Z-boson scalar and
vector potentials, and the Z-boson electric and magnetic fields whose
sources are electron and neutrino distributions and currents. The Z-boson
Maxwell equations are solved for point sources representing classical
point-like electrons and neutrinos. The parity violation of the weak
interation is manifest in the structure of these solutions. As an
application of this model, the neutral current contribution to the muonium
hyperfine structure is computed using nonrelativistic perturbation theory.

\end{abstract}

\section{INTRODUCTION}

The electromagnetic force is known to be the result of the exchange of
photons between charged particles. Outside of the context of quantum field
theory, this interaction is described in terms of classical scalar and
vector potentials ($A^{0}$ and $\mathbf{A}$) with their electric and
magnetic fields ($\mathbf{E}$ and $\mathbf{B}$). This description of the
electromagnetic interaction is a very useful model and forms the foundation
of a large body of macroscopic, atomic, and condensed matter physics.

The electron also participates in weak interactions, both charged current
(W-boson exchange) and neutral current (Z-boson exchange). For neutral
current events, electrons (or neutrinos) remain as electrons (or neutrinos).
We may imagine that one particle is the source of neutral current electric
and magnetic fields and that the other particle feels the forces produced by
these fields, in complete analogy with ordinary electromagnetism. In the
charged current case, an initial electron state emerges as a final neutrino
state or \textit{vice-versa}. For this reason it is difficult to consider
such a picture for the charged current interaction. Charged current
interactions will not be considered in this paper, and hence, we will
develop an incomplete picture of the weak force. This picture will, however,
be useful when an interaction cannot proceed by W-boson exchange..

This paper begins with an analysis of the standard model Lagrangian for the
first generation of leptons (the electron and its neutrino). A
nonrelativistic weak-field Hamiltonian for the electron is developed which
allows us to compute the interaction energy of an electron in the presence
of a classical Z-boson field. The Maxwell equations for the Z-boson are then
developed. In the absence of sources, the Maxwell equations are identical to
those of ordinary electromagnetism but with a massive photon. The Maxwell
equation source terms are derived from the interaction energies for both
electron and neutrino sources.

The Maxwell equations derived here can be used to describe the (albeit
small!) Z-boson field generated by macroscopic or atomic-scale distributions
of electrons. They may also be used to visualize the Z-boson fields
surrounding classical point-like electrons and neutrinos. The classical
point particle solutions provide an interesting visualization of the parity
violation in the standard model in terms of a vortex-like magnetic field
structure oriented with the electron's spin.

As an application of the formalism developed in this paper, a calculation of
the neutral current contribution to the hyperfine structure of muonium ($\mu
^{+}e^{-}$-bound state) is made using nonrelativistic perturbation theory.
Other applications might be found in nuclear physics.

\section{THE\ STANDARD\ MODEL\ LAGRANGIAN}

We begin by considering the matter terms of the Lagrangian density for the
standard model (with massless neutrinos) after spontaneous symmetry breaking
(following the notation of Quigg\cite{Quigg}):

\begin{eqnarray}
\mathcal{L}_{electron} &=&\overline{e}_{{\small R}}(i\gamma ^{\mu }\partial
_{\mu }+e\gamma ^{\mu }A_{\mu }-g^{\prime }\sin \theta _{{\small W}}\ \gamma
^{\mu }Z_{\mu })e_{{\small R}}  \nonumber \\
&&+\overline{e}_{{\small L}}(i\gamma ^{\mu }\partial _{\mu }+e\gamma ^{\mu
}A_{\mu }+\frac{1}{2}(g\cos \theta _{{\small W}}-g^{\prime }\sin \theta _{%
{\small W}})\gamma ^{\mu }Z_{\mu })e_{{\small L}}  \nonumber \\
&&-m\overline{e}_{{\small R}}e_{{\small L}}-m\overline{e}_{{\small L}}e_{%
{\small R}}\ ,  \nonumber \\
&& \\
\mathcal{L}_{neutrino} &=&\overline{\nu }(i\gamma ^{\mu }\partial _{\mu }-%
\frac{1}{2}(g\cos \theta _{{\small W}}+g^{\prime }\sin \theta _{{\small W}%
})\gamma ^{\mu }Z_{\mu })\nu \ .  \nonumber
\end{eqnarray}
Note that the $W_{\mu }^{\pm }$ boson fields have been set to zero,
decoupling electron and neutrino in the matter Lagrangian density. We do not
wish to consider charged current interactions. In addition, the neutrino
chirality is constrained such that $\frac{1}{2}(1+\gamma ^{5})\nu =0$.

The matter bispinors naturally decompose into left and right handed
chiralities when using the chiral representation\cite{Itzykson&Zuber} of the
Dirac matrices:

\begin{equation}
\gamma ^{\mu }=\left( 
\begin{array}{ll}
\left( 
\begin{array}{ll}
0 & -1 \\ 
-1 & 0
\end{array}
\right) & \left( 
\begin{array}{ll}
0 & \mathbf{\sigma } \\ 
-\mathbf{\sigma } & 0
\end{array}
\right)
\end{array}
\right) \ .
\end{equation}
In the chiral representation, the matter equations of motion are

\begin{eqnarray}
i\partial _{0}\left( 
\begin{array}{l}
e_{{\small R}} \\ 
e_{{\small L}}
\end{array}
\right)  &=&\mathcal{H}_{electron}\left( 
\begin{array}{l}
e_{{\small R}} \\ 
e_{{\small L}}
\end{array}
\right)   \label{EHam} \\
&=&\left( 
\begin{array}{ll}
\mathbf{\sigma }\cdot \mathbf{\pi }_{{\small R}}-\Phi _{{\small R}} & -m \\ 
-m & -\mathbf{\sigma }\cdot \mathbf{\pi }_{{\small L}}-\Phi _{{\small L}}
\end{array}
\right) \left( 
\begin{array}{l}
e_{{\small R}} \\ 
e_{{\small L}}
\end{array}
\right) \ ,  \nonumber \\
&&  \nonumber \\
i\partial _{0}\nu  &=&\mathcal{H}_{neutrino}\nu =\left( -\mathbf{\sigma }%
\cdot \mathbf{\pi }_{{\small n}}-\Phi _{{\small n}}\right) \nu \ ,
\label{NHam}
\end{eqnarray}
where

\begin{eqnarray}
\mathbf{\pi }_{{\small L}} &=&-i\mathbf{\nabla }+e\mathbf{A+}\frac{1}{2}%
(g\cos \theta _{{\small W}}-g^{\prime }\sin \theta _{{\small W}})\mathbf{Z\ ,%
}  \nonumber \\
\mathbf{\pi }_{{\small R}} &=&-i\mathbf{\nabla }+e\mathbf{A-}g^{\prime }\sin
\theta _{{\small W}}\ \mathbf{Z\ ,}  \nonumber \\
\Phi _{{\small L}} &=&eA^{0}+\frac{1}{2}(g\cos \theta _{{\small W}%
}-g^{\prime }\sin \theta _{{\small W}})Z^{0}\ ,  \nonumber \\
\Phi _{{\small R}} &=&eA^{0}\mathbf{-}g^{\prime }\sin \theta _{{\small W}}\
Z^{0}\ ,  \nonumber \\
\mathbf{\pi }_{{\small n}} &=&-i\mathbf{\nabla }-\frac{1}{2}(g\cos \theta _{%
{\small W}}+g^{\prime }\sin \theta _{{\small W}})\mathbf{Z\ ,} \\
\Phi _{{\small n}} &=&-\frac{1}{2}(g\cos \theta _{{\small W}}+g^{\prime
}\sin \theta _{{\small W}})Z^{0}\ ,  \nonumber \\
A^{\mu } &=&\left( 
\begin{array}{ll}
A^{0} & \mathbf{A}
\end{array}
\right) \ ,\ Z^{\mu }=\left( 
\begin{array}{ll}
Z^{0} & \mathbf{Z}
\end{array}
\right) \ .  \nonumber
\end{eqnarray}

\section{NONRELATIVISTIC-WEAK\ FIELD\ APPROXIMATION}

To assess the nonrelativistic limit it will be more convenient to use the
Dirac representation\cite{Itzykson&Zuber} for the Dirac Matrices:

\begin{equation}
\gamma ^{\mu }=\left( 
\begin{array}{ll}
\left( 
\begin{array}{ll}
1 & 0 \\ 
0 & -1
\end{array}
\right) & \left( 
\begin{array}{ll}
0 & \mathbf{\sigma } \\ 
-\mathbf{\sigma } & 0
\end{array}
\right)
\end{array}
\right) \ .
\end{equation}
In the chiral representation, the electron Hamiltonian in (\ref{EHam})
becomes

\begin{equation}
\mathcal{H}_{electron}=\left( 
\begin{array}{ll}
\begin{array}{l}
\frac{1}{2}\mathbf{\sigma }\cdot \left( \mathbf{\pi }_{{\small R}}-\mathbf{%
\pi }_{{\small L}}\right) \\ 
-\frac{1}{2}\left( \Phi _{{\small R}}+\Phi _{{\small L}}\right) +m \\ 
\ 
\end{array}
& 
\begin{array}{l}
\frac{1}{2}\mathbf{\sigma }\cdot \left( \mathbf{\pi }_{{\small R}}+\mathbf{%
\pi }_{{\small L}}\right) \\ 
-\frac{1}{2}\left( \Phi _{{\small R}}-\Phi _{{\small L}}\right) \\ 
\ 
\end{array}
\\ 
\begin{array}{l}
\frac{1}{2}\mathbf{\sigma }\cdot \left( \mathbf{\pi }_{{\small R}}+\mathbf{%
\pi }_{{\small L}}\right) \\ 
-\frac{1}{2}\left( \Phi _{{\small R}}-\Phi _{{\small L}}\right)
\end{array}
& 
\begin{array}{l}
\frac{1}{2}\mathbf{\sigma }\cdot \left( \mathbf{\pi }_{{\small R}}-\mathbf{%
\pi }_{{\small L}}\right) \\ 
-\frac{1}{2}\left( \Phi _{{\small R}}+\Phi _{{\small L}}\right) -m
\end{array}
\end{array}
\right) \ .
\end{equation}

In the nonrelativistic limit the electron bispinor may be written in the form

\begin{equation}
e=\left( 
\begin{array}{l}
\phi (\mathbf{x}) \\ 
\chi (\mathbf{x})
\end{array}
\right) e^{-i(m+E_{NR})t}\ ,
\end{equation}
where the nonrelativistic energy $E_{NR}\ll m$. We may solve for the
``small'' component spinor $\chi $ in terms of the ``large'' component
spinor $\phi $; whence, to lowest order in the ratio $E_{NR}/m$:

\begin{equation}
\chi =\frac{1}{2m}\left( \frac{1}{2}\mathbf{\sigma }\cdot \left( \mathbf{\pi 
}_{{\small R}}+\mathbf{\pi }_{{\small L}}\right) -\frac{1}{2}\left( \Phi _{%
{\small R}}-\Phi _{{\small L}}\right) \right) \phi \ .
\end{equation}
Substituting for $\chi $ in the equations of motion allows the
identification of the nonrelativistic Hamiltonian

\begin{eqnarray}
\mathcal{H}_{NR} &=&\frac{1}{2m}\left( \frac{1}{2}\mathbf{\sigma }\cdot
\left( \mathbf{\pi }_{{\small R}}+\mathbf{\pi }_{{\small L}}\right) -\frac{1%
}{2}\left( \Phi _{{\small R}}-\Phi _{{\small L}}\right) \right) ^{2} \\
&&+\frac{1}{2}\mathbf{\sigma }\cdot \left( \mathbf{\pi }_{{\small R}}-%
\mathbf{\pi }_{{\small L}}\right) -\frac{1}{2}\left( \Phi _{{\small R}}+\Phi
_{{\small L}}\right) \ .  \nonumber
\end{eqnarray}
Use the identity $\left( \mathbf{\sigma \cdot A}\right) \left( \mathbf{%
\sigma \cdot B}\right) =\mathbf{A\cdot B+}i\mathbf{\sigma \cdot }\left( 
\mathbf{A\times B}\right) $ and neglect terms quadratic in the fields (weak
field approximation)\ to derive

\begin{eqnarray}
\mathcal{H}_{NR} &=&-\frac{1}{2m}\mathbf{\nabla }^{2}+\mathcal{V\ }, \\
&&  \nonumber \\
\mathcal{V} &=&-\frac{i}{2m}\mathbf{\nabla \cdot }\left( e\mathbf{A}+\frac{1%
}{4}(g\cos \theta _{{\small W}}-3g^{\prime }\sin \theta _{{\small W}})%
\mathbf{Z}\right)   \nonumber \\
&&-\frac{i}{2m}\left( e\mathbf{A}+\frac{1}{4}(g\cos \theta _{{\small W}%
}-3g^{\prime }\sin \theta _{{\small W}})\mathbf{Z}\right) \cdot \mathbf{%
\nabla }  \nonumber \\
&&+\frac{1}{2m}\mathbf{\sigma }\cdot \left( \mathbf{\nabla \times }\left( e%
\mathbf{A}+\frac{1}{4}(g\cos \theta _{{\small W}}-3g^{\prime }\sin \theta _{%
{\small W}})\mathbf{Z}\right) \right)   \label{IntEn} \\
&&-i\frac{g\cos \theta _{{\small W}}+g^{\prime }\sin \theta _{{\small W}}}{8m%
}\left( \mathbf{\sigma }\cdot \mathbf{\nabla }Z^{0}+Z^{0}\mathbf{\sigma }%
\cdot \mathbf{\nabla }\right)   \nonumber \\
&&-eA^{0}-\frac{1}{4}(g\cos \theta _{{\small W}}-3g^{\prime }\sin \theta _{%
{\small W}})Z^{0}-\frac{1}{4}\left( g\cos \theta _{{\small W}}+g^{\prime
}\sin \theta _{{\small W}}\right) \mathbf{\sigma }\cdot \mathbf{Z\ .} 
\nonumber
\end{eqnarray}
Note that it is understood that the differential operators operate only on
the object immediately to its right.

\section{INTERACTION\ ENERGY}

The interaction energy (\ref{IntEn}) is responsible for the Maxwell equation
source terms. In the nonrelativistic-weak field approximation the
interaction energy for an electron source is just the expectation value of $%
\mathcal{V}$ using the nonrelativistic electron wavefunction $\phi $. After
integration by parts we have

\begin{eqnarray}
V &=&-\frac{i}{2m}\int \left( e\mathbf{A}+\frac{1}{4}(g\cos \theta _{{\small %
W}}-3g^{\prime }\sin \theta _{{\small W}})\mathbf{Z}\right) \cdot \left(
\phi ^{\dagger }\mathbf{\nabla }\phi -\left( \mathbf{\nabla }\phi ^{\dagger
}\right) \phi \right) d^{3}\mathbf{r}  \nonumber \\
&&+\frac{1}{2m}\int \left( \mathbf{\nabla \times }\left( e\mathbf{A}+\frac{1%
}{4}(g\cos \theta _{{\small W}}-3g^{\prime }\sin \theta _{{\small W}})%
\mathbf{Z}\right) \right) \cdot \left( \phi ^{\dagger }\mathbf{\sigma }\phi
\right) d^{3}\mathbf{r}  \nonumber \\
&&-i\frac{g\cos \theta _{{\small W}}+g^{\prime }\sin \theta _{{\small W}}}{8m%
}\int Z^{0}\left( \phi ^{\dagger }\mathbf{\sigma \cdot \nabla }\phi -\left( 
\mathbf{\nabla }\phi ^{\dagger }\right) \cdot \mathbf{\sigma }\phi \right)
d^{3}\mathbf{r}  \label{IntE} \\
&&-\int \left( eA^{0}+\frac{1}{4}(g\cos \theta _{{\small W}}-3g^{\prime
}\sin \theta _{{\small W}})Z^{0}\right) \phi ^{\dagger }\phi d^{3}\mathbf{r}
\nonumber \\
&&-\frac{1}{4}\left( g\cos \theta _{{\small W}}+g^{\prime }\sin \theta _{%
{\small W}}\right) \int \mathbf{Z\cdot }\left( \phi ^{\dagger }\mathbf{%
\sigma }\phi \right) d^{3}\mathbf{r\ .}  \nonumber
\end{eqnarray}

The nonrelativistic limit makes no sense for a massless neutrino. The
neutrino source interaction energy may be found directly from $\mathcal{H}%
_{neutrino}$ given by (\ref{NHam}):

\begin{eqnarray}
V &=&\frac{1}{2}\left( g\cos \theta _{{\small W}}+g^{\prime }\sin \theta _{%
{\small W}}\right) \int Z^{0}\nu ^{\dagger }\nu d^{3}\mathbf{r} \\
&&\mathbf{+}\frac{1}{2}\left( g\cos \theta _{{\small W}}+g^{\prime }\sin
\theta _{{\small W}}\right) \int \mathbf{Z\cdot }\left( \nu ^{\dagger }%
\mathbf{\sigma }\nu \right) d^{3}\mathbf{r\ .}  \nonumber
\end{eqnarray}

\section{MAXWELL\ EQUATIONS}

The first order Lagrangian for the photon is

\begin{equation}
L=\int \left( -\mathbf{E}\cdot \left( \mathbf{\nabla }A^{0}+\mathbf{\dot{A}}%
\right) -\mathbf{B\cdot }\left( \mathbf{\nabla \times A}\right) +\frac{1}{2}%
\left( |\mathbf{B}|^{2}-|\mathbf{E}|^{2}\right) \right) d^{3}\mathbf{r-}V\ .
\end{equation}
From this Lagrangian we derive the relationship between the potentials and
field strengths and two of the Maxwell equations: 
\begin{equation}
\mathbf{E}=\mathbf{-\nabla }A^{0}-\mathbf{\dot{A}\ ,\ \ B}=\mathbf{\nabla
\times A\ ,}
\end{equation}
\begin{equation}
\mathbf{\nabla \cdot E}=-e\phi ^{\dagger }\phi \ ,
\end{equation}
\begin{equation}
\mathbf{\nabla \times }\left( \mathbf{B+}\frac{e}{m}\frac{1}{2}\left( \phi
^{\dagger }\mathbf{\sigma }\phi \right) \right) -\mathbf{\dot{E}}=-e\left( 
\frac{-i}{2m}\right) \left( \phi ^{\dagger }\mathbf{\nabla }\phi -\left( 
\mathbf{\nabla }\phi ^{\dagger }\right) \phi \right) \ .  \label{PhotMax}
\end{equation}

The first order Lagrangian for the Z-boson field neglecting three and four
boson couplings (see Appendix A\ for further discussion)\ is

\begin{eqnarray}
L &=&\int \left( -\mathbf{E}\cdot \left( \mathbf{\nabla }Z^{0}+\mathbf{\dot{Z%
}}\right) -\mathbf{B\cdot }\left( \mathbf{\nabla \times Z}\right) +\frac{1}{2%
}\left( |\mathbf{B}|^{2}-|\mathbf{E}|^{2}\right) \right) d^{3}\mathbf{r} 
\nonumber \\
&&+\frac{1}{2}M^{2}\int \left( \left( Z^{0}\right) ^{2}-|\mathbf{Z}%
|^{2}\right) d^{3}\mathbf{r-}V\ .
\end{eqnarray}
This differs from the photon's Lagrangian by the addition of a mass term ($%
M=M_{Z}$). Note that the symbols ``$\mathbf{E}$'' and ``$\mathbf{B}$'' are
now used as the field strengths for the Z-boson instead of the photon. The
electron source terms are modified for the Z-boson:

\begin{eqnarray}
\mathbf{\nabla \cdot E+}M^{2}Z^{0} &=&-\frac{1}{4}(g\cos \theta _{{\small W}%
}-3g^{\prime }\sin \theta _{{\small W}})\phi ^{\dagger }\phi  \label{ZMax1}
\\
&&+\frac{1}{2}\left( g\cos \theta _{{\small W}}+g^{\prime }\sin \theta _{%
{\small W}}\right) \left( \frac{-i}{4m}\right) \left( \phi ^{\dagger }%
\mathbf{\sigma \cdot \nabla }\phi -\left( \mathbf{\nabla }\phi ^{\dagger
}\right) \cdot \mathbf{\sigma }\phi \right) \ ,  \nonumber
\end{eqnarray}

\begin{eqnarray}
&&\mathbf{\nabla \times }\left( \mathbf{B+}\frac{1}{4m}\left( g\cos \theta _{%
{\small W}}-3g^{\prime }\sin \theta _{{\small W}}\right) \frac{1}{2}\left(
\phi ^{\dagger }\mathbf{\sigma }\phi \right) \right) +M^{2}\mathbf{Z}-%
\mathbf{\dot{E}}  \nonumber \\
&=&-\frac{1}{4}(g\cos \theta _{{\small W}}-3g^{\prime }\sin \theta _{{\small %
W}})\left( \frac{-i}{2m}\right) \left( \phi ^{\dagger }\mathbf{\nabla }\phi
-\left( \mathbf{\nabla }\phi ^{\dagger }\right) \phi \right)  \label{ZMax2}
\\
&&+\frac{1}{2}\left( g\cos \theta _{{\small W}}+g^{\prime }\sin \theta _{%
{\small W}}\right) \frac{1}{2}\left( \phi ^{\dagger }\mathbf{\sigma }\phi
\right) \ .  \nonumber
\end{eqnarray}

Similar Maxwell equations for a massless neutrino are

\begin{equation}
\mathbf{\nabla \cdot E+}M^{2}Z^{0}=\frac{1}{2}\left( g\cos \theta _{{\small W%
}}+g^{\prime }\sin \theta _{{\small W}}\right) \nu ^{\dagger }\nu \ ,
\end{equation}

\begin{equation}
\mathbf{\nabla \times B}+M^{2}\mathbf{Z}-\mathbf{\dot{E}=-}\left( g\cos
\theta _{{\small W}}+g^{\prime }\sin \theta _{{\small W}}\right) \frac{1}{2}%
\left( \nu ^{\dagger }\mathbf{\sigma }\nu \right) \ .
\end{equation}

The remaining two Maxwell equations are just Bianchi identities and remain
unchanged from ordinary electromagnetism (Appendix A):

\begin{equation}
\mathbf{\nabla \cdot B}=0\ ,\ \ \mathbf{\nabla \times E}+\mathbf{\dot{B}}=0\
.
\end{equation}

\section{ELECTRON\ FIELDS}

It is interesting to consider the fields of a classical point-like electron.
First consider the photon fields. The wavefunction bilinear combinations
appearing as sources for the photon electric and magnetic fields have the
following interpretations:

\begin{eqnarray}
-e\phi ^{\dagger }\phi &\longrightarrow &\text{charge density }\rho \ , 
\nonumber \\
-e\left( \frac{-i}{2m}\right) \left( \phi ^{\dagger }\mathbf{\nabla }\phi
-\left( \mathbf{\nabla }\phi ^{\dagger }\right) \phi \right)
&\longrightarrow &\text{current density }\mathbf{j\ ,} \\
\frac{1}{2}\left( \phi ^{\dagger }\mathbf{\sigma }\phi \right)
&\longrightarrow &\text{spin angular momentum density }\mathbf{s\ .} 
\nonumber
\end{eqnarray}
The term $-\frac{e}{m}\frac{1}{2}\phi ^{\dagger }\mathbf{\sigma }\phi $ in (%
\ref{PhotMax}) has the physical interpretation of a magnetization (or a
dipole moment density). From these considerations we find that the
point-like classical electron has a charge -e and a magnetic moment of $-%
\frac{e}{m}$ times its spin (a gyromagnetic ratio of 2).

Now consider the Z-boson fields. There is the addition bilinear form $\left( 
\frac{-i}{4m}\right) \left( \phi ^{\dagger }\mathbf{\sigma \cdot \nabla }%
\phi -\left( \mathbf{\nabla }\phi ^{\dagger }\right) \cdot \mathbf{\sigma }%
\phi \right) $ which may be interpreted as a density for the helicity-%
\TEXTsymbol{\vert}velocity\TEXTsymbol{\vert} product but can be considered
vanishing for the electron at rest. The Z-boson Maxwell equations for the
point-like electron at rest with spin polarization ($S=\pm \frac{1}{2}$)
along the z-axis derived from (\ref{ZMax1}) and (\ref{ZMax2}) are then

\begin{equation}
\mathbf{\nabla \cdot E+}M^{2}Z^{0}=-\frac{1}{4}(g\cos \theta _{{\small W}%
}-3g^{\prime }\sin \theta _{{\small W}})\delta \left( \mathbf{r}\right) \ ,
\label{EMax1}
\end{equation}

\begin{eqnarray}
&&\mathbf{\nabla \times }\left( \mathbf{B+}\frac{1}{4m}\left( g\cos \theta _{%
{\small W}}-3g^{\prime }\sin \theta _{{\small W}}\right) S\widehat{\mathbf{z}%
}\delta \left( \mathbf{r}\right) \right) +M^{2}\mathbf{Z}  \label{EMax2} \\
&=&\frac{1}{2}\left( g\cos \theta _{{\small W}}+g^{\prime }\sin \theta _{%
{\small W}}\right) S\widehat{\mathbf{z}}\delta \left( \mathbf{r}\right) \ . 
\nonumber
\end{eqnarray}
Unlike the electromagnetic Maxwell equations for the electron at rest, there
is a nonvanishing magnetic source current density in (\ref{EMax2}) as well
as the familiar electric charge density in (\ref{EMax1}).

When the electron is placed in motion along the z-axis in either a state of
positive or negative helicity, additional source terms are required to
maintain the covariance of the Maxwell equations. Specifically, a magnetic
source current is generated due to the motion of the electric charge density
just as with ordinary electromagnetism; however, unlike ordinary
electromagnetism, an additional electric charge density coupled to the
nonvanishing at-rest magnetic source current is required for covariance.
Thus, in the limit of a classical point-like electron, we must make the
identification

\begin{equation}
\left( \frac{-i}{4m}\right) \left( \phi ^{\dagger }\mathbf{\sigma \cdot
\nabla }\phi -\left( \mathbf{\nabla }\phi ^{\dagger }\right) \cdot \mathbf{%
\sigma }\phi \right) \longrightarrow S\widehat{\mathbf{z}}\cdot \mathbf{v}%
\delta \left( \mathbf{r-v}t\right) \ ,
\end{equation}
and the Maxwell equations (\ref{EMax1}) and (\ref{EMax2}) become more
generally:

\begin{eqnarray}
\mathbf{\nabla \cdot E+}M^{2}Z^{0} &=&-\frac{1}{4}(g\cos \theta _{{\small W}%
}-3g^{\prime }\sin \theta _{{\small W}})\delta \left( \mathbf{r-v}t\right) \\
&&+\frac{1}{2}\left( g\cos \theta _{{\small W}}+g^{\prime }\sin \theta _{%
{\small W}}\right) S\widehat{\mathbf{z}}\cdot \mathbf{v}\delta \left( 
\mathbf{r-v}t\right) \ ,  \nonumber
\end{eqnarray}

\begin{eqnarray}
&&\mathbf{\nabla \times }\left( \mathbf{B+}\frac{1}{4m}\left( g\cos \theta _{%
{\small W}}-3g^{\prime }\sin \theta _{{\small W}}\right) S\widehat{\mathbf{z}%
}\delta \left( \mathbf{r-v}t\right) \right) +M^{2}\mathbf{Z}  \nonumber \\
&=&\frac{1}{2}\left( g\cos \theta _{{\small W}}+g^{\prime }\sin \theta _{%
{\small W}}\right) S\widehat{\mathbf{z}}\delta \left( \mathbf{r-v}t\right) \\
&&-\frac{1}{4}(g\cos \theta _{{\small W}}-3g^{\prime }\sin \theta _{{\small W%
}})\mathbf{v}\delta \left( \mathbf{r-v}t\right) \ .  \nonumber
\end{eqnarray}

The Maxwell equations for the point-like electron at rest (\ref{EMax1}) and (%
\ref{EMax2}) are solved in Appendix B. Physically there are three
contributions to the Z-boson fields. The first component is in the form of a
Yukawa potential due to a neutral current electric charge $q=-\frac{1}{4}%
(g\cos \theta _{W}-3g^{\prime }\sin \theta _{W})$:

\begin{eqnarray}
Z_{Yukawa}^{0} &=&\frac{q}{4\pi }\frac{e^{-Mr}}{r}\ ,  \nonumber \\
&&  \label{Yukawa} \\
\mathbf{E}_{Yukawa} &=&\frac{q}{4\pi }\left( \frac{M}{r}+\frac{1}{r^{2}}%
\right) e^{-Mr}\widehat{\mathbf{r}}\ ,  \nonumber
\end{eqnarray}

The second component is due to a neutral current magnetic dipole moment $\mu
=\frac{q}{m}S$; in spherical coordinates:

\begin{equation}
\mathbf{Z}_{Dipole}=\frac{\mu }{4\pi }\left( \frac{M}{r}+\frac{1}{r^{2}}%
\right) e^{-Mr}\sin \theta \widehat{\mathbf{\phi }}\ ,  \label{DPole1}
\end{equation}

\begin{eqnarray}
\mathbf{B}_{Dipole} &=&\frac{\mu }{2\pi }\left( \frac{M}{r^{2}}+\frac{1}{%
r^{3}}\right) e^{-Mr}\cos \theta \widehat{\mathbf{r}}  \label{DPole2} \\
&&+\frac{\mu }{4\pi }\left( \frac{M^{2}}{r}+\frac{M}{r^{2}}+\frac{1}{r^{3}}%
\right) e^{-Mr}\sin \theta \widehat{\mathbf{\theta }}+\frac{2}{3}\mu 
\widehat{\mathbf{z}}\delta \left( \mathbf{r}\right) \ .  \nonumber
\end{eqnarray}
Note that (\ref{DPole2}) implies that the electron has a neutral current
gyromagnetic ratio identical (when quantum filed theory radiative
corrections are ignored) to the electromagnetic.

The third component is a vortex-like magnetic field near the equatorial
plane due to a magnetic vortex charge $\kappa S=\frac{1}{2}\left( g\cos
\theta _{W}+g^{\prime }\sin \theta _{W}\right) S$:

\begin{eqnarray}
\mathbf{Z}_{PV} &=&-\frac{\kappa S}{2\pi }\left( \frac{1}{Mr^{2}}+\frac{1}{%
M^{2}r^{3}}\right) e^{-Mr}\cos \theta \widehat{\mathbf{r}}  \nonumber \\
&&-\frac{\kappa S}{4\pi }\left( \frac{1}{r}+\frac{1}{Mr^{2}}+\frac{1}{%
M^{2}r^{3}}\right) e^{-Mr}\sin \theta \widehat{\mathbf{\theta }}
\label{VTex1} \\
&&+\frac{1}{3}\frac{\kappa S}{M^{2}}\widehat{\mathbf{z}}\delta (\mathbf{r})\
,  \nonumber
\end{eqnarray}

\begin{equation}
\mathbf{B}_{PV}=\frac{\kappa S}{4\pi }\left( \frac{M}{r}+\frac{1}{r^{2}}%
\right) e^{-Mr}\sin \theta \widehat{\mathbf{\phi }}\ .  \label{VTex2}
\end{equation}
Note that (\ref{VTex1}) is not finite in the limit $M\rightarrow 0$. This is
because there are no nonvanishing solutions to $\mathbf{B}=\mathbf{\nabla
\times Z}$ and $\mathbf{\nabla \times B}=0$ with the same symmetries as (\ref
{VTex1}) and (\ref{VTex2}). The fields (\ref{VTex1}) and (\ref{VTex2}) have
no direct analogy in ordinary electromagnetism, although the axial symmetry
of the magnetic field (\ref{VTex2}) is similar to that of a charge in motion.

The presence of the $\delta $-functions in (\ref{DPole2}) and (\ref{VTex1})
(discussed further in Appendix B)\ are required topologically to preserve
Stokes integral identities.

The parameter $\kappa $ is a direct measure of the parity violation in the
standard model since it is the difference between the left-handed and
right-handed electron coupling constants, while the charge $q$ is their
average. Under parity inversions $\phi \rightarrow \pi +\phi $, $\theta
\rightarrow \pi -\theta $, and $\widehat{\mathbf{r}}\rightarrow \widehat{%
\mathbf{r}}$, $\widehat{\mathbf{\theta }}\rightarrow -\widehat{\mathbf{%
\theta }}$, $\widehat{\mathbf{\phi }}\rightarrow \mathbf{\hat{\phi}}$. For
the Yukawa part (\ref{Yukawa}) both $Z^{0}$ and $\mathbf{E}$ are even under
parity inversion. For the dipole part (\ref{DPole1}) and (\ref{DPole2}) $%
\mathbf{Z}$ is even and $\mathbf{B}$ is odd; however, for (\ref{VTex1}) and (%
\ref{VTex2}) $\mathbf{Z}$ is odd and $\mathbf{B}$ is even, signaling parity
violation.

The sign of the charge $q$ may be freely chosen, using $\sin ^{2}\theta
_{W}=0.23$, we have $|e|=0.30$, $|q|=0.014$, and $\kappa =-0.36\ sgn(q)$.

\section{NEUTRAL CURRENT CONTRIBUTIONS TO THE MUONIUM HYPERFINE STRUCTURE}

The $\mu ^{+}e^{-}$-bound state (muonium) is an important system for high
precision studies of quantum electrodynamics (QED)\cite{QED}. Its spectrum
is qualitatively similar to that of Hydrogen. Using the formalism in this
paper, it is a simple exercise in nonrelativistic perturbation theory to
compute the energy level shifts of muonium due to the neutral current
interaction. In particular, we calculate the neutral current perturbation to
the ground state energy of muonium, and the hyperfine splitting between the $%
J=0$ (singlet) and $J=1$ (triplet) ground states\cite{Muonium}.

Equation (\ref{IntE}) is the energy perturbation when $\phi \ (=\psi \times
spinor)$ is the electron's unperturbed nonrelativistic wave function, and $%
Z^{0}$ and $\mathbf{Z}$ are the potentials generated by a point-like
antimuon. The unperturbed ground state electron scalar wavefunction is

\begin{equation}
\psi =\frac{1}{\sqrt{\pi a^{3}}}e^{-r/a}\ ,
\end{equation}
where $a=4\pi /(m_{e}^{\prime }e^{2})$ is the muonium Bohr radius with $%
m_{e}^{\prime }$ being the reduced mass of the electron in muonium. The
potentials generated by the antimuon are given by equations (\ref{Yukawa})
through (\ref{VTex2}) with $q$ replaced by $-q$, $\kappa $ replaced by $%
-\kappa $, and the electron's mass replaced by the muon's. Since both
particles' spins may be freely oriented, it is necessary to replace the
fixed choices for the polarization axes with Pauli matrices, and in the end,
take the expectation value using a spin state describing both particles.
Following this procedure, there are three nonvanishing terms in (\ref{IntE}):

\begin{equation}
V_{C}=q\int Z^{0}\psi ^{\dagger }\psi d^{3}\mathbf{r=-}\frac{q^{2}}{\pi a^{3}%
}\frac{1}{\left( M+2/a\right) ^{2}}\ ,  \label{A}
\end{equation}

\begin{eqnarray}
V_{NPV} &=&-\frac{q}{2m}\int <\mathbf{B}\cdot \mathbf{\sigma }>\psi
^{\dagger }\psi d^{3}\mathbf{r}  \label{B} \\
&=&\frac{q^{2}}{m_{e}m_{\mu }}<\mathbf{S}_{e}\cdot \mathbf{S}_{\mu }>\frac{2%
}{3\pi a^{3}}\left( 1-\frac{M^{2}}{\left( M+2/a\right) ^{2}}\right) \ , 
\nonumber
\end{eqnarray}

\begin{eqnarray}
V_{PV} &=&-\frac{1}{2}\kappa \int <\mathbf{Z}\cdot \mathbf{\sigma }>\psi
^{\dagger }\psi d^{3}\mathbf{r}  \label{C} \\
&=&\frac{\kappa ^{2}}{3\pi a^{3}}<\mathbf{S}_{e}\cdot \mathbf{S}_{\mu
}>\left( \frac{2}{\left( M+2/a\right) ^{2}}+\frac{1}{M^{2}}\right) \ . 
\nonumber
\end{eqnarray}

The ``cross-terms'' in (\ref{IntE}) (proportional to $\kappa q$) involve the
expectation value $<\mathbf{S}_{e}\times \mathbf{S}_{\mu }>$ which can be
shown to be vanishing.

The expectation value of the electron and muon spin operator dot product $<%
\mathbf{S}_{e}\cdot \mathbf{S}_{\mu }>$ is evaluated by considering the
expectation of the total spin-squared $\mathbf{S}^{2}=(\mathbf{S}_{e}+%
\mathbf{S}_{\mu })^{2}$, using the decomposition of the symmetric triplet
spin states, and the antisymmetric singlet spin state into individual
particle spinors: $\uparrow \uparrow $, $\left( \uparrow \downarrow
+\downarrow \uparrow \right) /\sqrt{2}$, $\downarrow \downarrow $ (triplet),
and $\left( \uparrow \downarrow -\downarrow \uparrow \right) /\sqrt{2}$
(singlet). In the triplet state $<\mathbf{S}_{e}\cdot \mathbf{S}_{\mu }>=%
\frac{1}{4}$, and in the singlet state $<\mathbf{S}_{e}\cdot \mathbf{S}_{\mu
}>=-\frac{3}{4}$; hence, we may compute the neutral current contributions to
the hyperfine splitting of the ground state of 1.7 Hz from (\ref{B}), and 65
Hz from (\ref{C}), for a total of 67 Hz. This compares to a 4.5 Ghz
electromagnetic contribution which may be estimated from (\ref{B}) with $q$
replaced by $e$ and $M=0$. The contribution of (\ref{A}) shifts both singlet
and triplet ground states by -0.10 Hz.

The neutral current contributions to the spectrum of hydrogen\cite{Hydrogen}
could be analyzed in similar fashion with suitable generalization to include
the neutral current coupling of the proton. This coupling may be identified
by summing the contributions from the constituent quarks with inclusion of
radiative corrections\cite{quark}. Extension to include heavier elements is
limited by the complexity of atomic wavefunctions. However, for certain
cases, such as Cesium, accurate calculations can be performed\cite{Cesium}.
Also of interest are muonic atoms\cite{muonic} because the muon's Bohr
radius is much smaller than the electron's. In all these systems, the
neutral current interaction introduces parity violation into the atomic
structure due to the mixing of parity-even and parity-odd pure electronic
states.

The analysis of positronium ($e^{+}e^{-}$-bound state)\ is , however, more
problematic because of the annihilation process $e^{+}e^{-}\rightarrow \ $%
virtual photon $\rightarrow e^{+}e^{-}$. A precise treatment of positronium%
\cite{Karplus} requires using the Bethe-Salpeter equation for relativistic
bound states.

\section{NEUTRINO FIELDS}

The Maxwell equations for a point-like massless neutrino travelling through
the origin at $t=0$ towards the positive z-axis, after making the
identifications $\nu ^{\dagger }\nu \rightarrow \delta (\mathbf{r}-\widehat{%
\mathbf{z}}t)$ and $\frac{1}{2}\left( \nu ^{\dagger }\mathbf{\sigma }\nu
\right) \rightarrow S\widehat{\mathbf{z}}\delta (\mathbf{r}-\widehat{\mathbf{%
z}}t)$, are

\begin{equation}
\mathbf{\nabla \cdot E+}M^{2}Z^{0}=\kappa \delta (\mathbf{r}-\widehat{%
\mathbf{z}}t)\ ,  \label{NMAX1}
\end{equation}

\begin{equation}
\mathbf{\nabla \times B}+M^{2}\mathbf{Z}-\mathbf{\dot{E}=-}2\kappa S\widehat{%
\mathbf{z}}\delta (\mathbf{r}-\widehat{\mathbf{z}}t)\ .  \label{NMAX2}
\end{equation}

Appendix C considers the solution of the Maxwell equations (\ref{NMAX1}) and
(\ref{NMAX2}). A physical ($S=-\frac{1}{2}$) point-like massless neutrino
travelling along the z-axis at the speed of light generates scalar and
vector potentials nonvanishing only in the plane perpendicular to $\widehat{%
\mathbf{z}}$ at the neutrino's instantaneous position. An observer could
then only detect the neutrino's presence at its moment of closest approach.
This is not a property unique to the neutral current force; rather, it is a
property related to Lorentz transformations properties of potentials and
field strengths. The same phenomenon may be found in the ultra-relativistic
limit of the Lienard-Wiechart potential for the electron\cite{P&P}. The
neutral current fields are given in cylindrical coordinates below: 
\begin{eqnarray}
Z^{0} &=&\frac{\kappa }{2\pi }K_{0}(M\rho )\delta (z-t)\ ,\ \ \mathbf{Z}=%
\frac{\kappa }{2\pi }K_{0}(M\rho )\delta (z-t)\widehat{\mathbf{z}}\ , 
\nonumber \\
&&  \label{Nfields} \\
\mathbf{E} &=&\frac{\kappa M}{2\pi }K_{1}(M\rho )\delta (z-t)\widehat{%
\mathbf{\rho }}\ ,\ \ \mathbf{B}=\frac{\kappa M}{2\pi }K_{1}(M\rho )\delta
(z-t)\widehat{\mathbf{\phi }}\ .  \nonumber
\end{eqnarray}
where $\kappa $ is the neutrino's neutral current charge, and $K_{0}$ and $%
K_{1}$ are modified Bessel functions.

Under parity inversion the fields (\ref{Nfields}) represent a solution
propagating backwards in time (antiparticle). Under both parity and
time-reversal, $\mathbf{E}$ and $\mathbf{B}$ are even, while $Z^{0}$ and $%
\mathbf{Z}$ are odd.

If we consider solutions to the equations (\ref{NMAX1}) and (\ref{NMAX2})
for neutrinos of the wrong helicity ($S=+\frac{1}{2}$), the formal solution
to (\ref{NMAX1}) and (\ref{NMAX2}) would then contain delta function
derivatives (Appendix C).

\section{CONCLUSIONS}

A nonrelativistic weak-field Hamiltonian suitable for describing neutral
current interactions has been derived from the standard model. Maxwell
equations for the Z-boson fields have also been derived. They are suitable
for a description of the small Z-boson fields at an atomic or macroscopic
scale.

The fields surrounding classical point-like leptons due to the neutral
current interaction have been derived from the Z-boson Maxwell equations.
The electron possesses Z-boson electric and magnetic fields analogous to its
electromagnetic Coulomb and magnetic dipole fields. It also exhibits a
Z-boson magnetic field forming a vortex-like structure oriented with the
spin. This field is an explicit manifestation of the parity violation of the
standard model. The fields of a point-like neutrino have also been shown to
be nonvanishing only in the plane normal to its spin and containing the
particle.

Using the nonrelativistic Hamiltonian and the solution of the Maxwell
equations, we have calculated the 67 Hz contribution to the hyperfine
splitting of the muonium ground state using simple nonrelativistic
perturbation theory.

\section{ACKNOWLEDGMENT}

The author would like to thank Sean Ahern and John Norbury of the University
of Wisconsin at Milwaukee for their interest and encouragement.

\section{APPENDIX A: FIELD EQUATIONS FOR THE STANDARD MODEL}

The gauge terms in the standard model Lagrangian are

\[
\mathcal{L}_{gauge}=-\frac{1}{4}F_{a\mu \nu }F_{a}^{\mu \nu }-\frac{1}{4}%
f_{\mu \nu }f^{\mu \nu }\ , 
\]
\begin{equation}
F_{a\mu \nu }=\partial _{\nu }b_{a\mu }-\partial _{\mu }b_{a\nu
}+g\varepsilon _{abc}b_{b\mu }b_{c\nu }\ ,  \tag{A1}
\end{equation}
\[
f_{\mu \nu }=\partial _{\nu }\mathcal{A}_{\mu }-\partial _{\mu }\mathcal{A}%
_{\nu }\ . 
\]
where $\mathcal{A}_{\mu }$ and $b_{a\mu }$ are respectively the U(1)\ and
SU(2) gauge fields, linear combinations of which form the physical boson
fields:

\[
W^{\pm }=\left( b_{1}\mp ib_{2}\right) /\sqrt{2}\ , 
\]
\begin{equation}
Z=-\mathcal{A}\sin \theta _{W}+b_{3}\cos \theta _{W}\ ,  \tag{A2}
\label{Z&A}
\end{equation}
\[
A=\mathcal{A}\cos \theta _{W}+b_{3}\sin \theta _{W}\ . 
\]
Using the definitions (A2), field strengths for the physical fields may be
defined as

\[
F_{\mu \nu }^{\pm }=\partial _{\nu }W_{\mu }^{\pm }-\partial _{\mu }W_{\nu
}^{\pm }\pm ig\cos \theta _{W}\left( W_{\mu }^{\pm }Z_{\nu }-Z_{\mu }W_{\nu
}^{\pm }\right) \pm ie\left( W_{\mu }^{\pm }A_{\nu }-A_{\mu }W_{\nu }^{\pm
}\right) \ , 
\]
\begin{equation}
F_{\mu \nu }^{Z}=\partial _{\nu }Z_{\mu }-\partial _{\mu }Z_{\nu }-ig\cos
\theta _{W}\left( W_{\mu }^{+}W_{\nu }^{-}-W_{\mu }^{-}W_{\nu }^{+}\right) \
,  \tag{A3}
\end{equation}
\[
F_{\mu \nu }^{A}=\partial _{\nu }A_{\mu }-\partial _{\mu }A_{\nu }-ie\left(
W_{\mu }^{+}W_{\nu }^{-}-W_{\mu }^{-}W_{\nu }^{+}\right) \ . 
\]
where $e=g\sin \theta _{W}=g^{\prime }\cos \theta _{W}$. The gauge
Lagrangian density, now including the boson mass terms, becomes

\begin{equation}
\mathcal{L}_{gauge}=-\frac{1}{4}F_{\mu \nu }^{A}F^{A\mu \nu }-\frac{1}{4}%
F_{\mu \nu }^{Z}F^{Z\mu \nu }-\frac{1}{2}F_{\mu \nu }^{+}F^{-\mu \nu
}+M_{W}^{2}W_{\mu }^{+}W^{-\mu }+\frac{1}{2}M_{Z}^{2}Z_{\mu }Z^{\mu }\ . 
\tag{A4}
\end{equation}

The equations of motion are

\begin{eqnarray*}
&&\partial ^{\nu }F_{\mu \nu }^{\pm }\pm ig\cos \theta _{W}F_{\mu \nu }^{\pm
}Z^{\nu }\pm ieF_{\mu \nu }^{\pm }A^{\nu }\mp ieF_{\mu \nu }^{A}W^{\pm \nu
}\mp ig\cos \theta _{W}F_{\mu \nu }^{Z}W^{\pm \nu }+M_{W}^{2}W_{\mu }^{+} \\
&=&\text{ weak charged current source (\textit{e.g.}, }e\rightarrow \nu 
\text{ and }\nu \rightarrow e\text{)\ ,}
\end{eqnarray*}
\begin{eqnarray}
&&\partial ^{\nu }F_{\mu \nu }^{Z}+ig\cos \theta _{W}F_{\mu \nu }^{-}W^{+\nu
}-ig\cos \theta _{W}F_{\mu \nu }^{+}W^{-\nu }+M_{Z}^{2}Z_{\mu }  \tag{A5}
\label{FS} \\
&=&\text{ weak neutral current source (\textit{e.g.}, }e\rightarrow e\text{
and }\nu \rightarrow \nu \text{)\ ,}  \nonumber
\end{eqnarray}
\begin{eqnarray*}
&&\partial ^{\nu }F_{\mu \nu }^{A}+ieF_{\mu \nu }^{-}W^{+\nu }-ieF_{\mu \nu
}^{+}W^{-\nu } \\
&=&\text{electromagnetic source (\textit{e.g.}, }e\rightarrow e\text{)\ .}
\end{eqnarray*}
where the matter terms in the full Lagrangian produce the sources for the
boson fields in equations (A5).

To characterize the classical fields surrounding a lepton we consider only
configurations for which the lepton remains invariant in its initial state.
There are no weak charged currents and hence $W^{\pm }=0$ is a consistent
solution to the fields equations and the remaining fields equations for $A$
and $Z$ decouple.

There are also Bianchi identities for the fields strengths which are
satisfied identically:

\[
\partial ^{\nu }\ ^{*}F_{\mu \nu }^{\pm }\pm ig\cos \theta _{W}\ ^{*}F_{\mu
\nu }^{\pm }Z^{\nu }\pm ie^{*}F_{\mu \nu }^{\pm }A^{\nu }\mp ie\ ^{*}F_{\mu
\nu }^{A}W^{\pm \nu }\mp ig\cos \theta _{W}\ ^{*}F_{\mu \nu }^{Z}W^{\pm \nu
}=0\ , 
\]
\newline
\begin{equation}
\partial ^{\nu }\ ^{*}F_{\mu \nu }^{Z}+ig\cos \theta _{W}\ ^{*}F_{\mu \nu
}^{-}W^{+\nu }-ig\cos \theta _{W}\ ^{*}F_{\mu \nu }^{+}W^{-\nu }=0\ , 
\tag{A6}
\end{equation}
\[
\partial ^{\nu }\ ^{*}F_{\mu \nu }^{A}+ie\ ^{*}F_{\mu \nu }^{-}W^{+\nu }-ie\
^{*}F_{\mu \nu }^{+}W^{-\nu }=0\ , 
\]
where $\ ^{*}F^{\mu \nu }=\frac{1}{2}\varepsilon ^{\mu \nu \rho \sigma
}F_{\rho \sigma }$. When $W^{\pm }=0$ the Bianchi identities decouple, and
since there are no boson mass terms in the Bianchi identities, they reduce
to the familiar form of classical electrodynamics for both the photon and
the Z-boson.

The structure of the classical field theories for $A$ and $Z$ are now
identical except that the Z-boson has a mass $M=M_{Z}$ which appears in its
equation of motion but not in its Bianchi identity.

\section{APPENDIX B: SOLUTION FOR THE ELECTRON}

Without sources, $\mathbf{B}$ and $\mathbf{Z}$ are solutions to $\mathbf{%
\nabla \times }\left( \mathbf{\nabla \times U}\right) +M^{2}\mathbf{U}=0$.
When $\mathbf{U}$ is a solution then so is $\mathbf{\nabla \times U}$. Two
solution pairs in spherical coordinates are

\begin{equation}
\mathbf{U=}z^{-\frac{1}{2}}\QATOPD\{ \} {I_{\frac{3}{2}}(z)}{K_{\frac{3}{2}%
}(z)}\sin \theta \ \widehat{\mathbf{\phi }}\ ,  \tag{B1}
\end{equation}
\[
\mathbf{\nabla \times U=}2Mz^{-\frac{3}{2}}\QATOPD\{ \} {I_{\frac{3}{2}%
}(z)}{K_{\frac{3}{2}}(z)}\cos \theta \ \widehat{\mathbf{r}}-Mz^{-\frac{3}{2}%
}\QATOPD\{ \} {zI_{\frac{1}{2}}(z)-I_{\frac{3}{2}}(z)}{-zK_{\frac{1}{2}%
}(z)-K_{\frac{3}{2}}(z)}\sin \theta \ \widehat{\mathbf{\theta }}\ , 
\]

where $I_{n}(z)$ and $K_{n}(z)$ are modified Bessel functions with $z=Mr$;

explicitly

\[
I_{\frac{3}{2}}(z)=\sqrt{\frac{2}{\pi z}}\left( \cosh z-\frac{\sinh z}{z}%
\right) \ ,\ \ I_{\frac{1}{2}}(z)=\sqrt{\frac{2}{\pi z}}\sinh z\ , 
\]
\begin{equation}
K_{\frac{3}{2}}(z)=\sqrt{\frac{\pi }{2z}}\left( 1+\frac{1}{z}\right) e^{-z}\
,\ \ K_{\frac{1}{2}}(z)=\sqrt{\frac{\pi }{2z}}e^{-z}\ .  \tag{B2}
\end{equation}

Consider a spherical volume $V$ containing the uniform magnetic source
density $\frac{\varkappa }{V}S\widehat{\mathbf{z}}=\frac{\varkappa }{V}%
S\left( \cos \theta \ \widehat{\mathbf{r}}-\sin \theta \ \widehat{\mathbf{%
\theta }}\right) $. Consider homogeneous solutions for $\mathbf{Z}$\ and $%
\mathbf{B}$ inside (upper) and outside (lower) the sphere:

\[
\mathbf{Z}=2z^{-\frac{3}{2}}\QATOPD\{ \} {aI_{\frac{3}{2}}(z)}{bK_{\frac{3}{2%
}}(z)}\cos \theta \ \widehat{\mathbf{r}}-z^{-\frac{3}{2}}\QATOPD\{ \} {azI_{%
\frac{1}{2}}(z)-aI_{\frac{3}{2}}(z)}{-bzK_{\frac{1}{2}}(z)-bK_{\frac{3}{2}%
}(z)}\sin \theta \ \widehat{\mathbf{\theta }}\ , 
\]
\begin{equation}
\mathbf{B}=-Mz^{-\frac{1}{2}}\QATOPD\{ \} {aI_{\frac{3}{2}}(z)}{bK_{\frac{3}{%
2}}(z)}\sin \theta \ \widehat{\mathbf{\phi }}\ .  \tag{B3}
\end{equation}
A particular solution to $\mathbf{\nabla \times B}+M^{2}\mathbf{Z=}\frac{%
\varkappa }{V}S\widehat{\mathbf{z}}$ must be added inside the sphere, such as

\begin{equation}
\mathbf{Z}=M^{-2}\frac{\varkappa }{V}S\widehat{\mathbf{z}}\ ,\ \ \mathbf{B}%
=0\ .  \tag{B4}
\end{equation}

The constants $a$ and $b$ are determined from the boundary conditions that
the components of $\mathbf{Z}$ and $\mathbf{B}$ parallel to the interface at 
$r=R$ are continuous across that interface:

\begin{equation}
a=-\frac{3\kappa }{4\pi }SM\left( MR\right) ^{-\frac{3}{2}}K_{\frac{3}{2}%
}(MR)\rightarrow R^{-3}\text{ as }R\rightarrow 0\ ,  \tag{B5}
\end{equation}
\[
b=-\frac{3\kappa }{4\pi }SM\left( MR\right) ^{-\frac{3}{2}}I_{\frac{3}{2}%
}(MR)\rightarrow -\frac{\kappa }{4\pi }SM\sqrt{\frac{2}{\pi }}\text{ as }%
R\rightarrow 0\ . 
\]

The point-like electron solution corresponds to the $R\rightarrow 0$ limit;
however, note that the constant $a$ diverges like $R^{-3}$ implying that $%
\mathbf{Z}$ may possess a $\delta $-function at the origin. To determine its
weight, integrate $\mathbf{\nabla \times B}+M^{2}\mathbf{Z}$ over a
spherical volume and apply Stokes' theorem for volume integration:

\begin{equation}
\kappa S\widehat{\mathbf{z}}=\int (\mathbf{\nabla \times B})d^{3}\mathbf{r+}%
M^{2}\int \mathbf{Z}d^{3}\mathbf{r}=\int (d^{2}\mathbf{S\times B})\mathbf{+}%
M^{2}\int \mathbf{Z}d^{3}\mathbf{r\ }.  \tag{B6}
\end{equation}

This holds for any radius but in the limit as the radius tends to $\infty $
the magnetic field vanishes. $\mathbf{Z}$ must have a term $\frac{\kappa S}{%
3M^{2}}\delta \left( \mathbf{r}\right) \widehat{\mathbf{z}}$ for equality to
hold.

Next consider a spherical volume $V$ which contains the uniform
magnetization $\mathbf{M=}\frac{\mu }{V}\widehat{\mathbf{z}}=\frac{\mu }{V}%
\left( \cos \theta \ \widehat{\mathbf{r}}-\sin \theta \ \widehat{\mathbf{%
\theta }}\right) $. Solutions for $\mathbf{Z}$\ and $\mathbf{B}$ are

\begin{equation}
\mathbf{Z}=-z^{-\frac{1}{2}}\QATOPD\{ \} {aI_{\frac{3}{2}}(z)}{bK_{\frac{3}{2%
}}(z)}\sin \theta \ \widehat{\mathbf{\phi }}\ ,  \tag{B6}
\end{equation}
\[
\mathbf{B}=-2Mz^{-\frac{3}{2}}\QATOPD\{ \} {aI_{\frac{3}{2}}(z)}{bK_{\frac{3%
}{2}}(z)}\cos \theta \ \widehat{\mathbf{r}}+Mz^{-\frac{3}{2}}\QATOPD\{ \}
{azI_{\frac{1}{2}}(z)-aI_{\frac{3}{2}}(z)}{-bzK_{\frac{1}{2}}(z)-bK_{\frac{3%
}{2}}(z)}\sin \theta \ \widehat{\mathbf{\theta }}\ . 
\]
The vector potential $\mathbf{Z}$ must be continuous across the interface at 
$r=R$. The other boundary condition involves $\mathbf{H}=\mathbf{B}-\mathbf{M%
}$. It is easy to demonstrate using Stokes' theorem that, without surface
current sources, the tangential component of $\mathbf{H}$ must be continuous
across the interface, The solutions for the constants $a$ and $b$ are

\begin{equation}
a=-\frac{3}{4\pi }\mu M^{2}\left( MR\right) ^{-\frac{3}{2}}K_{\frac{3}{2}%
}(MR)\rightarrow R^{-3}\text{ as }R\rightarrow 0  \tag{B7}
\end{equation}
\[
b=-\frac{3}{4\pi }\mu M^{2}\left( MR\right) ^{-\frac{3}{2}}I_{\frac{3}{2}%
}(MR)\rightarrow -\frac{\mu }{4\pi }M^{2}\sqrt{\frac{2}{\pi }}\text{ as }%
R\rightarrow 0 
\]
The divergence of the constant $a$ signifies the presence of a $\delta $%
-function in the magnetic field. To determine its weight, compute the volume
integral of $\mathbf{\nabla \times Z=B}$ and use the Stokes theorem for
volume integration as before. There is a similar singularity in the
point-like electron's electromagnetic dipole field\cite{Jackson}.

The solution for the electric field is a straightforward application of
Gauss' law.

\section{APPENDIX C\ - SOLUTION FOR THE MASSLESS NEUTRINO}

Consider a massless neutrino whose trajectory takes it through the origin
and then along the positive z-axis. Its potentials and fields will be
functions of $u=z-t$ and the cylindrical coordinate $\rho $. The Maxwell
equations (\ref{NMAX1}) and (\ref{NMAX2}) may be written as second order
differential equations for the potentials using cylindrical coordinates:

\begin{equation}
-\frac{\partial ^{2}Z^{+}}{\partial \rho ^{2}}-\frac{1}{\rho }\frac{\partial
Z^{+}}{\partial \rho }+M^{2}Z^{+}-2\frac{\partial ^{2}Z^{-}}{\partial u^{2}}+%
\frac{\partial }{\partial u}\left( \frac{\partial Z^{\rho }}{\partial \rho }+%
\frac{1}{\rho }Z^{\rho }\right) =\frac{\kappa }{2}\left( 1-2S\right) \delta (%
\mathbf{r}-\widehat{\mathbf{z}}t)  \tag{C1}  \label{Z1}
\end{equation}
\begin{equation}
-\frac{\partial ^{2}Z^{-}}{\partial \rho ^{2}}-\frac{1}{\rho }\frac{\partial
Z^{-}}{\partial \rho }+M^{2}Z^{-}=\frac{\kappa }{2}\left( 1+2S\right) \delta
(\mathbf{r}-\widehat{\mathbf{z}}t)  \tag{C2}  \label{Z2}
\end{equation}
\begin{equation}
\frac{\partial ^{2}Z^{-}}{\partial \rho \partial u}-\frac{1}{2}Z^{\rho }=0 
\tag{C3}  \label{Z3}
\end{equation}
\begin{equation}
-\frac{\partial ^{2}Z^{\phi }}{\partial \rho ^{2}}-\frac{1}{\rho }\frac{%
\partial Z^{\phi }}{\partial \rho }+\left( \frac{1}{\rho ^{2}}+M^{2}\right)
Z^{\phi }=0  \tag{C4}  \label{Z4}
\end{equation}
where $Z^{+}=\frac{1}{2}\left( Z^{0}+\widehat{\mathbf{z}}\mathbf{\cdot Z}%
\right) $, $Z^{-}=\frac{1}{2}\left( Z^{0}-\widehat{\mathbf{z}}\mathbf{\cdot Z%
}\right) $, $Z^{\rho }=\widehat{\mathbf{\rho }}\cdot \mathbf{Z}$, and $%
Z^{\phi }=\widehat{\mathbf{\phi }}\cdot \mathbf{Z}$.

Equation (C4) is the modified Bessel equation. With no singularity at the
origin, equation (C4) has no bounded solution at infinity other than $%
Z^{\phi }=0$. Similarly, for $S=-\frac{1}{2}$ we must choose $Z^{-}=Z^{\rho
}=0$. Equation (C1) is then satisfied by the modified Bessel function $%
Z^{+}(\rho ,u)=aK_{0}(M\rho )\delta (u)$. To determine the weight $a$,
Gauss's law is applied to a tube of radius $R$ concentric with the z-axis:

\begin{eqnarray}
\kappa &=&\int \mathbf{E}\cdot d^{2}\mathbf{S+}M^{2}\int Z^{0}d^{3}\mathbf{r}%
=2\pi aMRK_{1}(MR)+2\pi aM^{2}\int_{0}^{R}K_{0}(M\rho )\rho d\rho  \nonumber
\\
&=&2\pi a\lim_{\rho \rightarrow 0}M\rho K_{1}(M\rho )=2\pi a  \tag{C5}
\label{C5}
\end{eqnarray}

When $S=+\frac{1}{2}$, equation (C2) has a solution proportional to $%
K_{0}(M\rho )\delta (u)$. The $\delta $-function is then differentiated in
(C1) and (C3).

\end{document}